\documentclass[runningheads]{llncs}
\usepackage{graphicx}
\usepackage{subfigure}
\usepackage{ulem}
\usepackage{colortbl}
\usepackage[section]{placeins}
\usepackage{float}
\usepackage{graphicx}
\usepackage{marvosym}
\begin{document}
\title{ECCR: Edge-Cloud Collaborative Recovery for Low-Power Wide-Area Networks interference mitigation}
\vspace{-0.8cm}

\author{Luoyu Mei\inst{1}\and
Zhimeng Yin\inst{2} \and
Xiaolei Zhou \inst{1,3}\textsuperscript{\Letter} \and
Shuai Wang\inst{1} \and
Kai Sun\inst{1}}
\institute{Southeast University, Nanjing, CHINA \\
\email{\{lymei2002, shuaiwang, sunk\}@seu.edu.cn}
\and City University of Hong Kong, Hong Kong, CHINA \\
\email{\{zhimeyin\}@cityu.edu.hk}
\and The Sixty-Third Research Institute, National University of Defense Technology, Changsha, CHINA\\
\email{\{zhouxiaolei\}@nudt.edu.cn}}

\authorrunning{Luoyu, et al.}
\titlerunning{ECCR: Edge-Cloud Collaborative Recovery}
\maketitle 

\vspace{-0.8cm}
\begin{abstract}
  Recent advances in Low-Power Wide-Area Networks have mitigated interference by using cloud assistance. Those methods transmit the RSSI samples and corrupted packets to the cloud to restore the correct message. However, the effectiveness of those methods is challenged by the high transmission data amount. This paper presents a novel method for interference mitigation in a Edge-Cloud collaborative manner, namely ECCR. It does not require  transmitting RSSI sample any more, whose length is eight times of the packet's. We demonstrate the disjointness of the bit errors of packets at the base stations via real-word experiments. ECCR leverages this to collaborate with multiple base stations for error recovery. Each base station detects and reports bit error locations to the cloud, then both error checking code and interfered packets from other receivers are utilized to restore correct packets. ECCR takes the advantages of both the global management ability of the cloud and the signal to perceive the benefit of each base station, and it is applicable to deployed LP-WAN systems (e.g. sx1280) without any extra hardware requirement. Experimental results show that ECCR is able to accurately decode packets when packets have nearly $51.76\%$ corruption.

\vspace{-0.3cm}
\end{abstract}

\vspace{-1.2cm}
\section{\uppercase{Introduction}}
\vspace{-0.3cm}

Low-Power Wide-Area Networks (LP-WANs) are gaining increasing attention in both the industry and academia, because of the advantages of long-range coverage, low energy consumption, and low deployment cost. Among LP-WANs technologies, LoRa is one of the leading emergent technologies in the unlicensed sub-GHz bands. It covers an area of several square kilometers from the base station and supports millions of end devices in the field
 \cite{ref7}. However, with the widely application of different wireless technologies in daily life and industry \cite{ref7,ref8,ref9,ref9,ref10,ref11,ref12}, multiple wireless protocols might be densely deployed in the same area, such as LoRa, sigfox \cite{ref28}, and 802.11ah \cite{ref8}. As a result, those wireless networks are inevitably overlapping, and lead to either isomorphism or heterogeneity interferences.

Most of the conventional approaches mitigate the LoRa interference by re-designing the Physical and MAC layers \cite{ref3,ref4,ref5,ref6}.
Transparent solutions are proposed in \cite{ref5,ref6} to re-design and synchronize LoRa sender, while \cite{ref1,ref6} take efforts to avoid corruptions at the base station side. 
Those efforts introduce extra hardware cost, deployment complexity, or are not compatible with deployed LoRa devices. Some recent efforts make use of the cloud resources to mitigate the LoRa interference, without any extra hardware. OPR \cite{ref2} restores the corrupted packets by transmitting those packets and RSSI samples to the cloud, and cycles through alternative fragments matched in the error-detection fields. However, those cloud-based methods lead to excessive overhead of RSSI transmission and computation, which greatly limits their feasibility in practice.

Inspired by the cloud-based methods, we ask a natural question that can we further reduce the transmission and computation overhead. In this paper, we propose a novel method for LoRa interference mitigation in an edge-cloud collaborative manner, named as Edge-Cloud Collaborative Recovery (ECCR). Instead of directly transmitting the RSSI samples, we identify the corruptions by adding error checking codes at the base station side. Besides, we find that bit errors of the packet received by different base stations are disjoint with others. In a nutshell, corruptions are detected by the error checking code before decoding. And then, the packets from multiple receivers can be utilized to restore the packet at the cloud side. Since the errors in packets are located on the base station, there is no need to transfer RSSI samples to the cloud any more. Benefit from this, ECCR greatly reduces both the transmission and computation overhead in the conventional cloud-based approach. 

To support edge-cloud collaboration, the challenge for ECCR in a LoRa base station is to rapidly detect packet corruptions quickly enough so that it ensures recovering packets in real-time communication. We design error checking code after the encoding of the LoRa physical payload, through which, the base station detects corruptions before decoding the packet. With such error checking code, ECCR successfully detects and reports corruption for using cloud resources to restore packets.

This paper presents the first edge-cloud collaborative design for LoRa interference mitigation. The features we provide and the challenges we address in this emulation-based design are indeed generic and applicable to a whole set of future interference mitigation. Specifically, the major contributions of ECCR are as follows:

\vspace{-0.3cm}
\begin{itemize}
    \item[$\bullet$] We propose ECCR, a novel interference mitigation approach for LoRa. To the best of our knowledge, it is the first edge-cloud collaborate method for interference mitigation. It takes the advantage of both clouds’ global management ability and edges’ signal perceive benefits. Without modifying any hardware, ECCR is a transparent design and hence is easily deployed in existing LoRa infrastructure.
    
    \item[$\bullet$] To mitigate interference in real-time, we address a few challenges including (i) detecting corruption before decoding the packets, (ii) collaborating multiple base stations for packet recovery. These techniques provide guidance for the range extension of edge-cloud collaborative interference mitigation.
    
    \item[$\bullet$] We conduct extensive experiments on commercial-off-the-shelf devices (sx1280 LoRa chip) and the USRP-N210 platform as the base station to evaluate the correctness and performance of ECCR. Experimental results show that ECCR is able to accurately decode packets even the corruption rate achieves $51.76\%$.
    
\end{itemize}
\vspace{-0.6cm}
\section{\uppercase{Background and motivation}}
\vspace{-0.3cm}

To explain ECCR, it is necessary to first introduce how LoRa and LoRa’s Wide-Area Networking architecture (LoRaWAN) work. In this section, we first concisely introduce the architecture of LoRa and  LoRa’s Wide-Area Networking architecture (LoRaWAN) and then explain the principles of ECCR, finally, we conduct experiments to motivate our works.

\vspace{-0.5cm}
\subsection{How LoRa Works}
\vspace{-0.3cm}

LoRa is a new type of spread spectrum communication protocol released by Semtech in August 2013. It works in the unlicensed sub-GHz frequency. As shown in Fig. 1, LoRa employs the technology of Chirp Spread Spectrum. Since the characteristics of long-range and high robustness, it has been utilized for decades in military communications. Recently, it has become the de-facto mainstream technology for the Internet of Things (IoT) for both industrial and civilian networks worldwide.

\vspace{-0.8cm}
\begin{figure}[H]
    \centering
    \includegraphics[width=4.5cm]{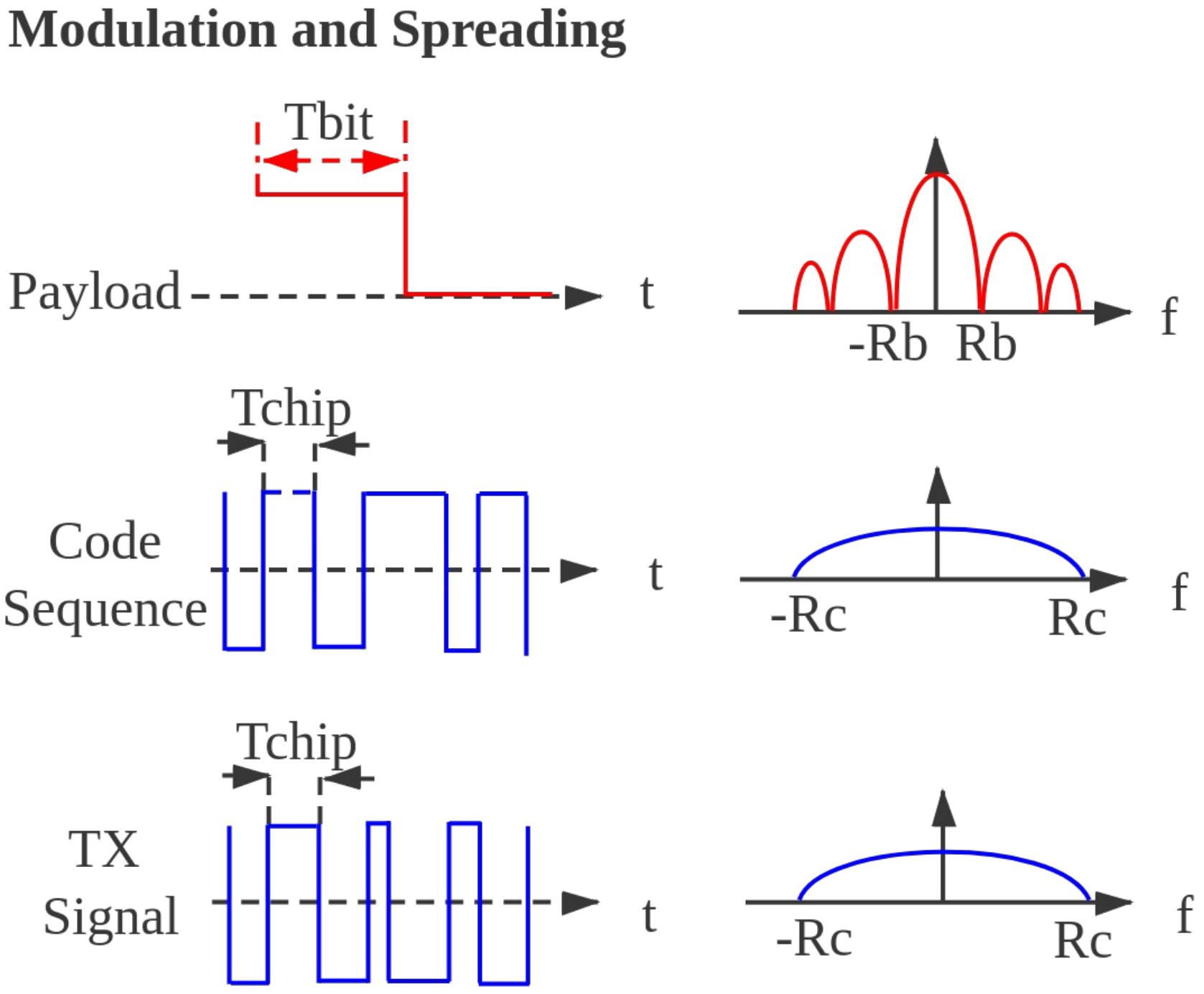}
    \caption{\textbf{Spread Spectrum of LoRa \cite{ref29}}}
    \label{Fig01}
    \vspace{-0.3cm}
\end{figure}
\vspace{-1.0cm}
\subsection{How LoRaWAN Works}
\vspace{-0.2cm}

The architecture of LoRaWAN is shown in Fig. 2. It contains several functional modules: Receiver, Gateway, Network service, Application. Generally, LoRa end node utilizes sub-GHz band wireless for data transmission with base stations, and after receiving the mixed signals, the gateway demodulates them into packets. These packets are finally transmitted to the cloud for applicational usage.    

\vspace{-0.8cm}
\begin{figure}[H]
    \centering
    \includegraphics[width=9cm]{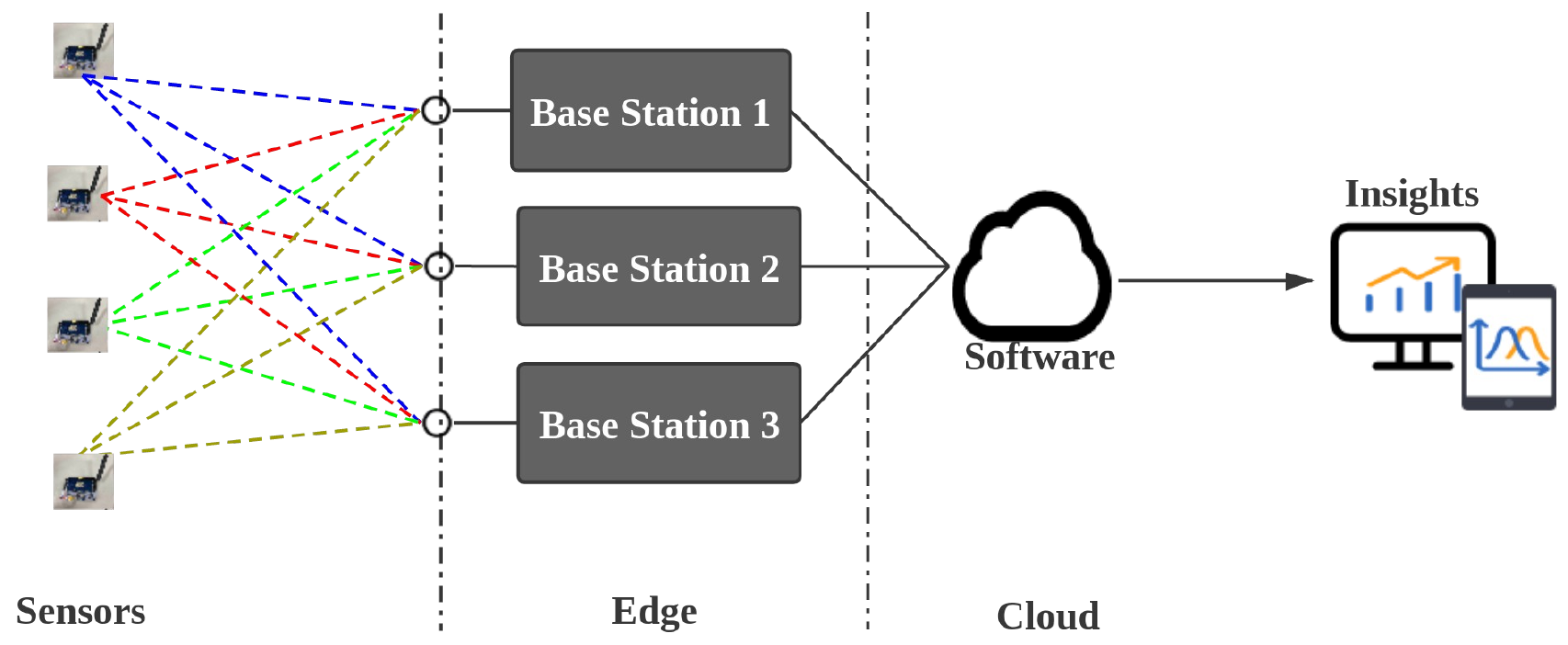}
    \caption{\textbf{LoRaWAN Architecture}}
    \label{Fig02}
\end{figure}
\vspace{-0.8cm}

\vspace{-0.9cm}
\begin{table}[H]
    \centering
    \begin{tabular}{| >{\centering} p{95 pt} | >{\centering} p{90 pt} | >{\centering} p{80 pt} | >{\centering} p{70 pt} |}
        \hline
        & \textbf{Data Transmission amount} & \textbf{Error correction capability} & \textbf{Computational 
complexity}\cr
        \hline
        \textbf{Standard} \cite{ref16} &	Low	 & Low	& Low \cr
        \hline
        \textbf{Cloud-based} \cite{ref2} &	High &	High &	High \cr
        \hline 
        \textbf{Edge-Cloud: ECCR} & 	\textbf{Low} &	\textbf{High} &	\textbf{Low}  \cr
        \hline
    \end{tabular}
    \caption{\textbf{Motivation of the ECCR}}
     \label{Table01}
\end{table}
\vspace{-0.9cm}
\vspace{-1.0cm}
\subsection{Motivations}

\vspace{-0.3cm}
Prior works has shown massive improvement in performance with modified hardware to coherently combine signals at the physical layer. However the extra hardware cost indeed limits the feasibility in real system. Balanuta et al. propose a cloud-based approach in  \cite{ref2} to leverage most likely corrupt bits across a set of packets that suffered failed CRCs at multiple LoRa LP-WAN basestations. After offloading the corrupted packets and RSSI samples, the failed packets might be recovered with a probability of 72\%. However, the offloading phase incurs excessive communication overhead and limits the large-scale application. We summarize the major performances of Standard LoRa[17], Conventional approach with specialized hardware, and  Cloud-based approach in Table 1. In this paper, we ask a natural question that can we design an approach that achieves all the ideal performance at the same time, i.e., low data transmission amount, low computational complexity, high error correction capability  and no extra hardware demand . To achieve this,  we design an interference  mitigation approach in a edge-cloud collaborative manner. The corruption is detected at base station side while the packets are restored at the cloud side. Such a design greatly reduces the data transmission amount. Besides, ECCR utilizes packets from multiple base stations, to achieve high error correction capability. With a carefully designed packet recovery algorithm, ECCR is able to restore packets with the time approaching LoRa.

\vspace{-0.9cm}
\begin{figure}[H]
    \centering
    \includegraphics[width=10cm]{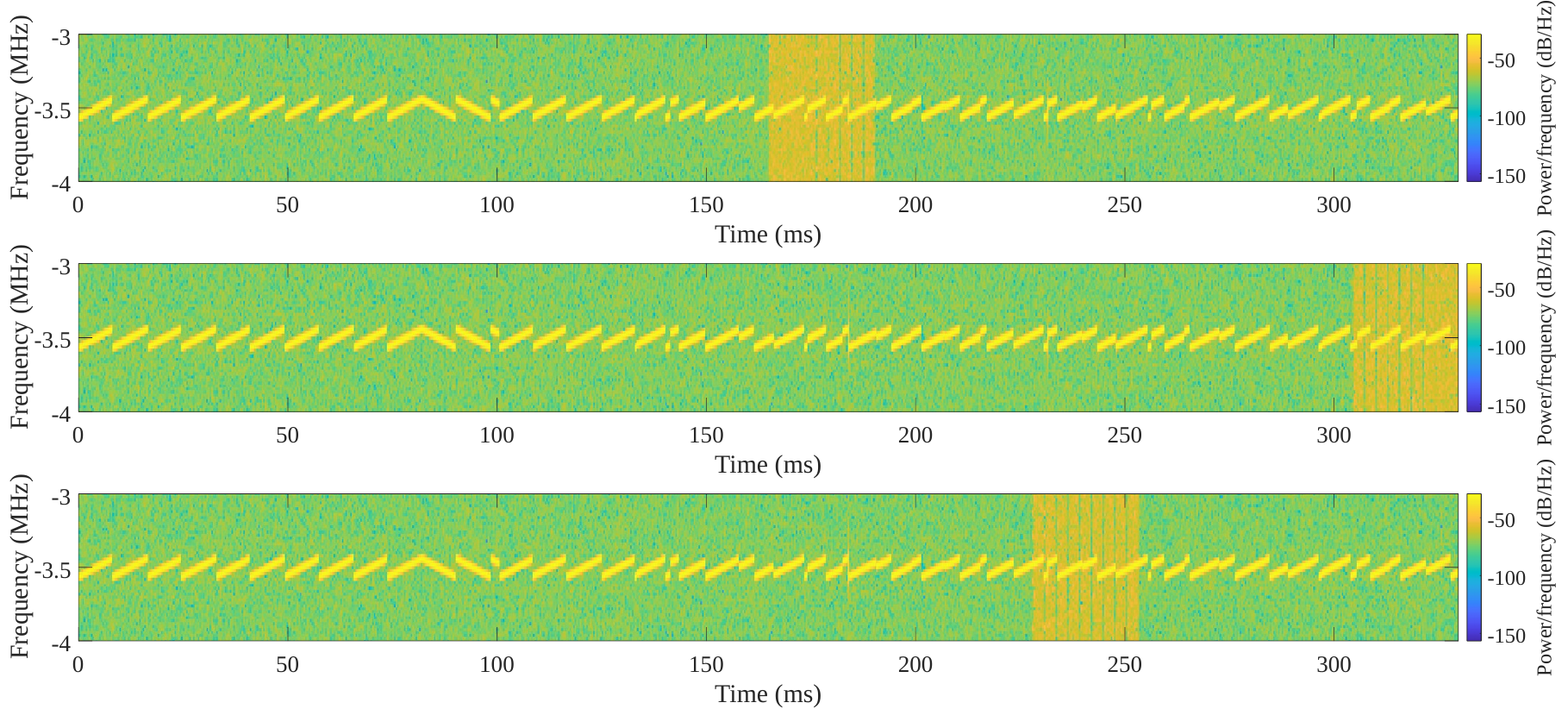}
    \caption{\textbf{In Phase/in Quadrature (I/Q) of a LoRa packet with Collision captured with a software defined Radio. (From top to bottom are Lab, Hallway, Library)}}
    \label{Fig03}
\end{figure}
\vspace{-0.8cm}

\vspace{-0.8cm}
\begin{table}[H]
    \centering
   
    \begin{tabular}{|p{50 pt}|p{250pt}|}
        \hline
        & \textbf{\centerline{Payload Received}} \cr
        \hline
        \textbf{Lab} & \textbf{\color{red}{74 86 111}} … 108 111 32 87\quad… 114 108 100 33 …  \cr
        \hline
        \textbf{Hallway} & 72 101 108 …  108 111 32 87\quad… \textbf{\color{red}{98 108 117 49}} …    \cr
        \hline
        \textbf{Library} & 72 101 108 … \textbf{\color{red}{105 119 32 78}}… 114 108 100 33…  \cr
        \hline
    \end{tabular}
    \caption{\textbf{Payload corruption in different receivers. (Bold part represent for corruptions)}.}
     \label{Table02}
\end{table}
\vspace{-0.8cm}

\vspace{-0.8cm}
\subsubsection{Disjoint Interference.} Our design is based on an interesting finding. For different wireless devices, their coverage vary a lot due to the difference in protocols.  We find that, the interference is disjoint among different LoRa base stations. To support this findings, we conduct a real experiment. Our first micro-benchmark shows the difference of interference in different receivers.

Fig. 3 shows the corruptions in different receivers, which are disjoint. Table 1 also shows that the received payloads of LoRa are corrupted at different locations when facing interference. 

We utilized a real-world, 10 sq. km. test-bed in Southeast University, to collect LoRa packets with interference. We examine the interference of LoRa in different sites: (i) a laboratory room, (ii) a hallway, and (iii) a library. We set the transmission power at 10dBm and put the sender in outdoor environments.

\vspace{-0.7cm}
\subsubsection{Benefit of Low Data Transmission Requirement.} Received Signal Strength Indication (RSSI) is an indication of the strength of the received signal, and its realization is carried out after the baseband receiving filter of the reverse channel. Traditionally, RSSI is utilized to determine the link quality and whether to increase the signal sending power.

OPR \cite{ref2} proposes a cloud-based error detecting method. It requires the base station sending RSSI samples as an index for detecting interference. Since, LoRa is able to work under the power level of the noise floor, burst interference increases RSSI level, and then corruptions can be identified in the cloud side. However, in LoRa protocols, RSSI is eight times of the length of the payload (e.g. 200 bytes for a 25-byte payload packet [2]). Compared to the payload, RSSI sample is still very long, even after compression. Clearly, transmitting the RSSI samples to the cloud leads to high the network throughputs of base stations.   

\vspace{-0.6cm}

\section{\uppercase{Main design}}
\vspace{-0.3cm}

ECCR takes advantages of both the global management ability of the cloud and the signal awareness benefit of each base station. In this section, we first describe the overview of ECCR, then move forward and step into the key components of ECCR, i.e., error detection and error recovery, repectively. 

\vspace{-0.8cm}
\begin{figure}[H]
    \centering
    \includegraphics[width=7cm]{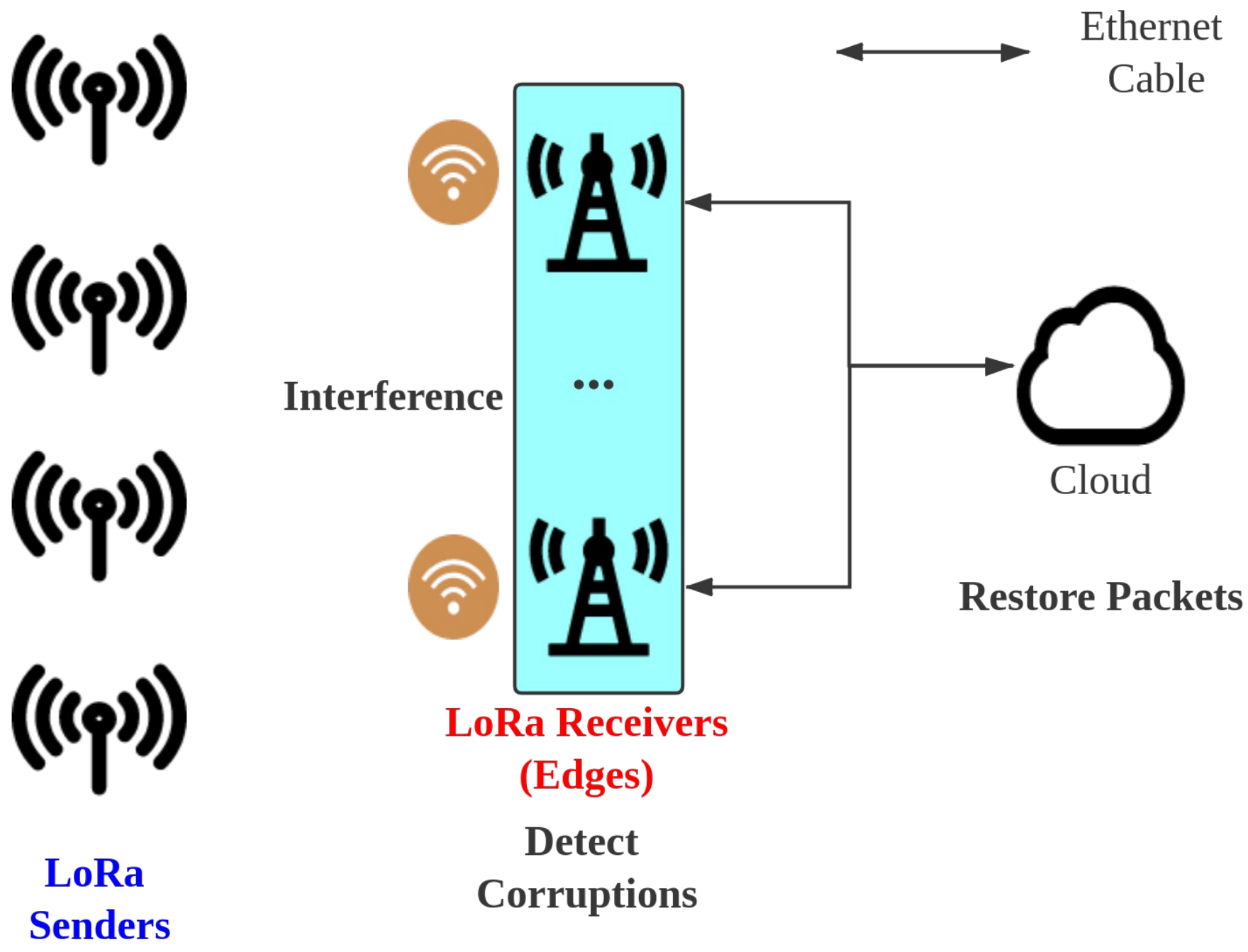}
    \caption{\textbf{The Architecture of the ECCR}}
    \label{Fig04}
\end{figure}
\vspace{-1.2cm}

\subsection{Overview of the ECCR}
\vspace{-0.3cm}

Fig. 4 shows the architecutre of ECCR.The LoRa senders send packets to the base sations in the field. ECCR adds error checking code after the encoding of LoRa physical payload for error detection, to detects corruption of the received packets in the base station and report them to the cloud. Since the corrupted parts of those packets are disjoint, when multiple packets form different base stations are available, the cloud restore the corrupted packets with a weight voting algorithm. Generally, the error correction capability of ECCR has a growth trend as the increasing number of base stations, because an increasing number of useful packets are collected by base stations. 

\vspace{-0.4cm}
\subsection{Error Detection}
\vspace{-0.3cm}

Error Detection is the most important part of the ECCR design. ECCR adds error checking codes in the LoRa physical payloads so that the base station can identify whether a received packet is corrupted by the interference before decoding.  

\vspace{-0.8cm}
\begin{figure}[H]
    \centering
    \includegraphics[width=9cm]{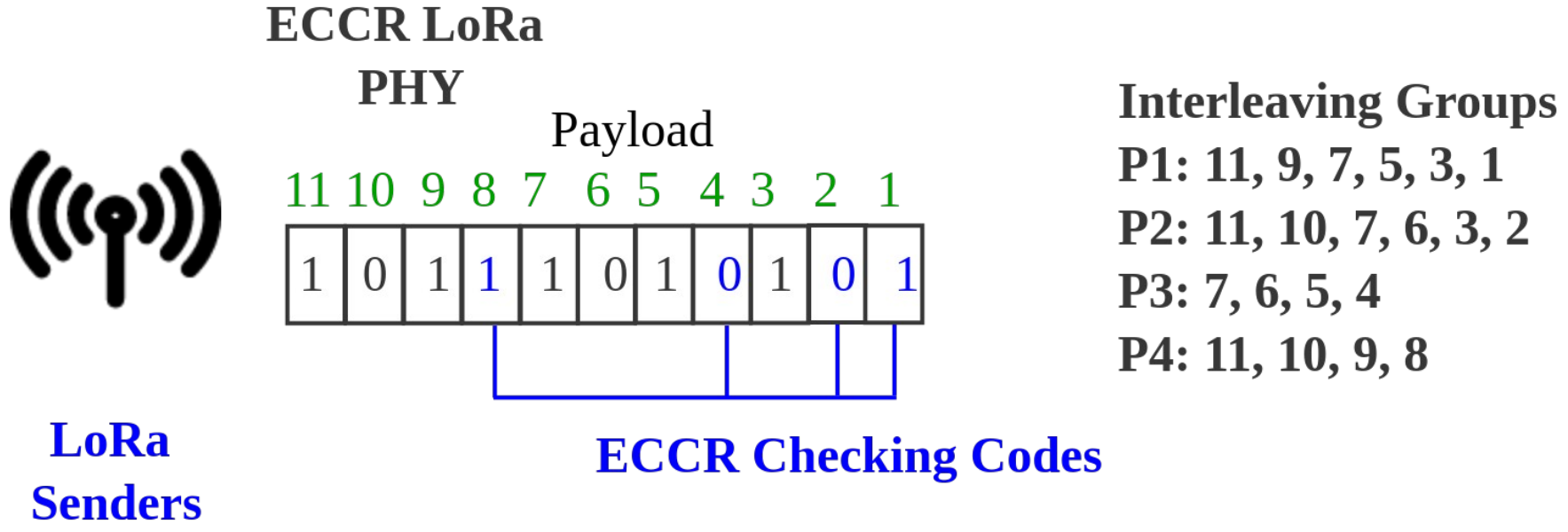}
    \caption{\textbf{How ECCR Works in LoRa Senders}}
    \label{Fig05}
\end{figure}
\vspace{-0.8cm}

Fig. 5 shows the ECCR checking codes after encoding a LoRa physical payload. We take the idea of hamming for error checking, and add checking bits into the payload. The number of checking bits in a packet is counted with the Equation: $2^r \geq m + r + 1$, where, $r$ is the number of checking bits, $m$ is data bits.

As shown in Fig.5. For example, when the data bits is 7, r equals to 4, since $2^4 \geq 7 + 4 + 1$. The checking codes only add 4 more bits into a 7-bit payload. Specifically, checking bits are located in the $2^k$-th bit in the new payload, where $k=1,2,...r$. In the above example, they are 1, 2, 4, 8 , repectively. Each checking bit represents for a interleave group (e.g. $G_1 - G_4$). Group $G_k$ indexes for the bits which are located in where the $k$-th bit of binary representation of the location is 1 (e.g.  $G_1$ index for 11, 9, 7, 5, 3, 1). 
Although the packet length will slightly increas(4 bits in 7-bit payload). Our approach avoids the transmission of RSSI samples (add 200 bytes to a 25-byte payload packet \cite{ref2}). Generally, ECCR reduces the computation overhead of error detection.

\vspace{-1.0cm}
\begin{figure}[H]
    \centering
    \includegraphics[width=12cm]{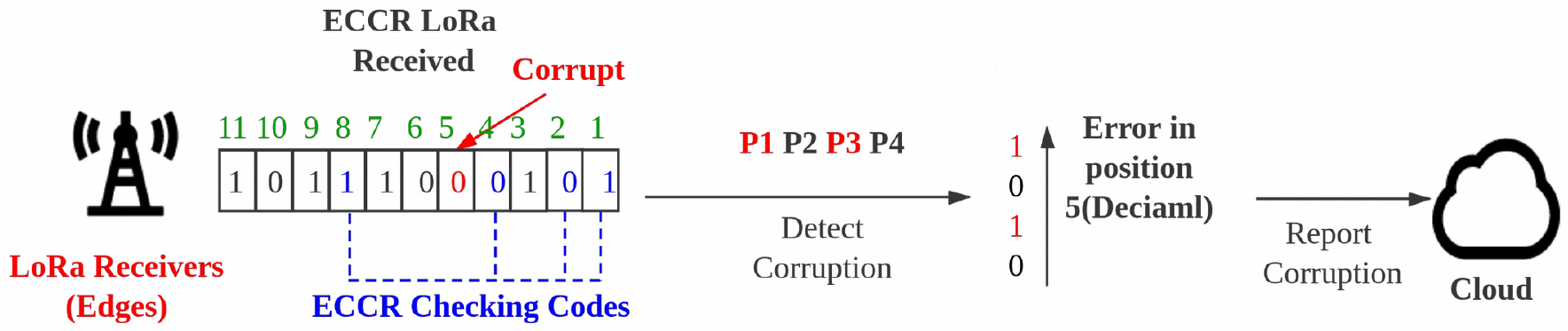}
    \caption{\textbf{How ECCR Carries on Error Detection in LoRa Receivers}}
    \label{Fig06}
\end{figure}
\vspace{-0.8cm}

Fig. 6 show the error detection works in the base stations. The ECCR checking codes are used for detecting corruptions before decoding packets. After demodulating the signal, the base station utilizes ECCR checking to detect corruption, before decoding. If the bit in location $k$ is wrong, the interleave groups with index it, will fail the error checking. In our design, "1" is used to represent for error, while "0" for correct, and the binary sequence of right and wrong cases of interleaving groups meets the error location when converted to decimal. ECCR takes advantage of this and detects error locations. For example, if the $5$-th bit corrupted, the correct and wrong situation of the four interleaving groups are as the equation (1):

\vspace{-0.3cm}
\begin{equation}
    \left\{
   \begin{array}{l}
    Gb_1 = D_1 \oplus  D_3 \oplus  D_5 \oplus  D_7 \oplus  D_9 \oplus \\
    Gb_2 = D_2 \oplus  D_3 \oplus  D_6 \oplus  D_7 \oplus  D_{10} \oplus D_{11} \\
    Gb_3 = D_4 \oplus  D_1 \oplus  D_6 \oplus  D_7  \\
    Gb_4 = D_8 \oplus  D_8 \oplus  D_9 \oplus  D_{10} \oplus  D_{11} 
	\end{array}
    \right.
\label{eq02}    
\end{equation}
\vspace{-0.3cm}

Here $D_k$ represents the correct and wrong conditions of $k$-th bit, and  $Gb_{1-4}$ represents the Boolean value of interleaving groups.

Once a corruption packet is detected, the base stations reports to the cloud. Then the cloud is able to collaborate packets from multiple base stations and restore packets through voting. Although ECCR checking code is able to correct some error bits, it has restrictions when corruptions increase (e.g. When 7, 11 corrupted, four interleave groups all come to failure, the error cannot be recovered by ECCR checking code), so that ECCR further utilizes the cloud to recovery packets.

\vspace{-0.3cm}
\vspace{-0.3cm}
\subsection{Error Recovery}
\vspace{-0.3cm}

We have demonstrated in Section 2.3 that the received payloads of LoRa are corrupted at a different location when facing disjoint interference in multiple base stations. ECCR further utilizes the error checking code to detect corruptions in packets, then the error is reported to the cloud through reliable ethernet connections during which LoRaWAN utilizes 128 bits AES for integrity protection and data encryption. The cloud collaborates packets from multiple base stations, assigns weight to them according to the proportion of corruption, and utilizes a weight voting algorithm to restore the correct packet. Specifically, the weight of symbols is signed according to the equation (2):  

\vspace{-0.3cm}
\begin{equation}
    \left\{
   \begin{array}{l}
    W_k = \frac{\sum_{i=1}^{L_k}k(2)[t] - \sum_{i=1}^{L_k}(k(2)[t] == 1 \land G_i) }{\sum_{i=1}^{L_k}k(2)[t]} \\
    L_k = \sum_{i=1}^{p}k(2)[i] \lor 1 (\exists k(2)[p] \land \xout{\exists} k(2)[p+1] )\\
    G_i = \land_{n=1}^{\cup(D  in  G_i)}D_n
	\end{array}
    \right.
\label{eq03}    
\end{equation}
\vspace{-0.3cm}

Where $W_k$ represents for the weight of the $k$-th symbol,  $k(2)$ for the binary representation of $k$, $L_k$ for the length of $k(2)$, and $G_i$ for the ECCR checking result for $i$-th interleaving group, $D_n$ for the correct and wrong conditions of the $n$-th bit. $D$ in $G_i$ are shown in equation (2).

\vspace{-0.8cm}
\begin{table}[H]
    \scriptsize
    \centering
    \begin{tabular}{| p{35 pt} | p{170pt} | p{130pt}| }
        \hline
        & \textbf{\centerline{Payload Received}} & \textbf{\centerline{Symbol Weights}} \cr
        \hline
        \textbf{Lab} & \textbf{74 86 111} … 108 111 32 87\quad… 114 108 100 33& \textbf{\color{red}{0 0 0}} … 100 0 50 33 … 100 50 50 33 \cr
        \hline
        \textbf{Hallway} & 72 101 108 …  108 111 32 87\quad… \textbf{98 108 117 49}& 0 0 0 … 100 0 50 33 … \textbf{\color{red}{ 0 0 0 0 }} \cr
        \hline
        \textbf{Library} & 72 101 108 … \textbf{105 119 32 78} … 114 108 100 33& 0 0 0 … \textbf{\color{red}{0 0 0 0}} … 100 50 50 33\cr
        \hline
        & \textbf{\centerline{Voting Result}}& \cr
        \hline
        \textbf{Voting} & 72 101 108 …  108 111 32 87\quad… 114 108 100 33& \cr
        \hline
    \end{tabular}
    \caption{\textbf{Voting Result (Bold part represent for corruptions).}}
     \label{Table03}
\end{table}
\vspace{-1.2cm}

ECCR utilizes the weight equation (2) to assign weights for each symbols. The weights are also utilized to measure the reliability of packets during the weight voting process. A higher weight means the packet is closer to the correct one. In that way, the correct information of multiple packets are collaborated to restore the ture packet (an example of weight voting process is shown in Table 3). Note that, if all the weights of the symbols in those packets are 0, in other words, interleaving groups that index the symbol location all come to error, ECCR treats every packet equally at that symbol location. (e.g. $1$, $2$ and $3$-th symbol in Table 3)

\vspace{-1.0cm}
\section{\uppercase{Performance evaluation}}
\vspace{-0.8cm}

To evaluate the correctness and performance of the proposed ECCR approach, we conduct extensive experiments with emulation. Our evaluations include different situations of Wi-Fi interference. To ensure limit-testing the performance of ECCR, we emulated In-Phase/Quadrature (I/Q) of LoRa packets using LoRaMatlab \cite{ref13} and utilize WLAN Waveform Generator \cite{ref14} to generate Wi-Fi packets as interference. We control the degree of interference by extending the time of Wi-Fi packets. Our test covers multiple scenarios, including: (i) Standard LoRa, (ii) LoRa and ECCR checking code, (iii) ECCR.     

\vspace{-0.8cm}
\begin{figure}[H]
    \begin{minipage}[t]{0.2\linewidth}
        \centering
        \includegraphics[width=2.3cm]{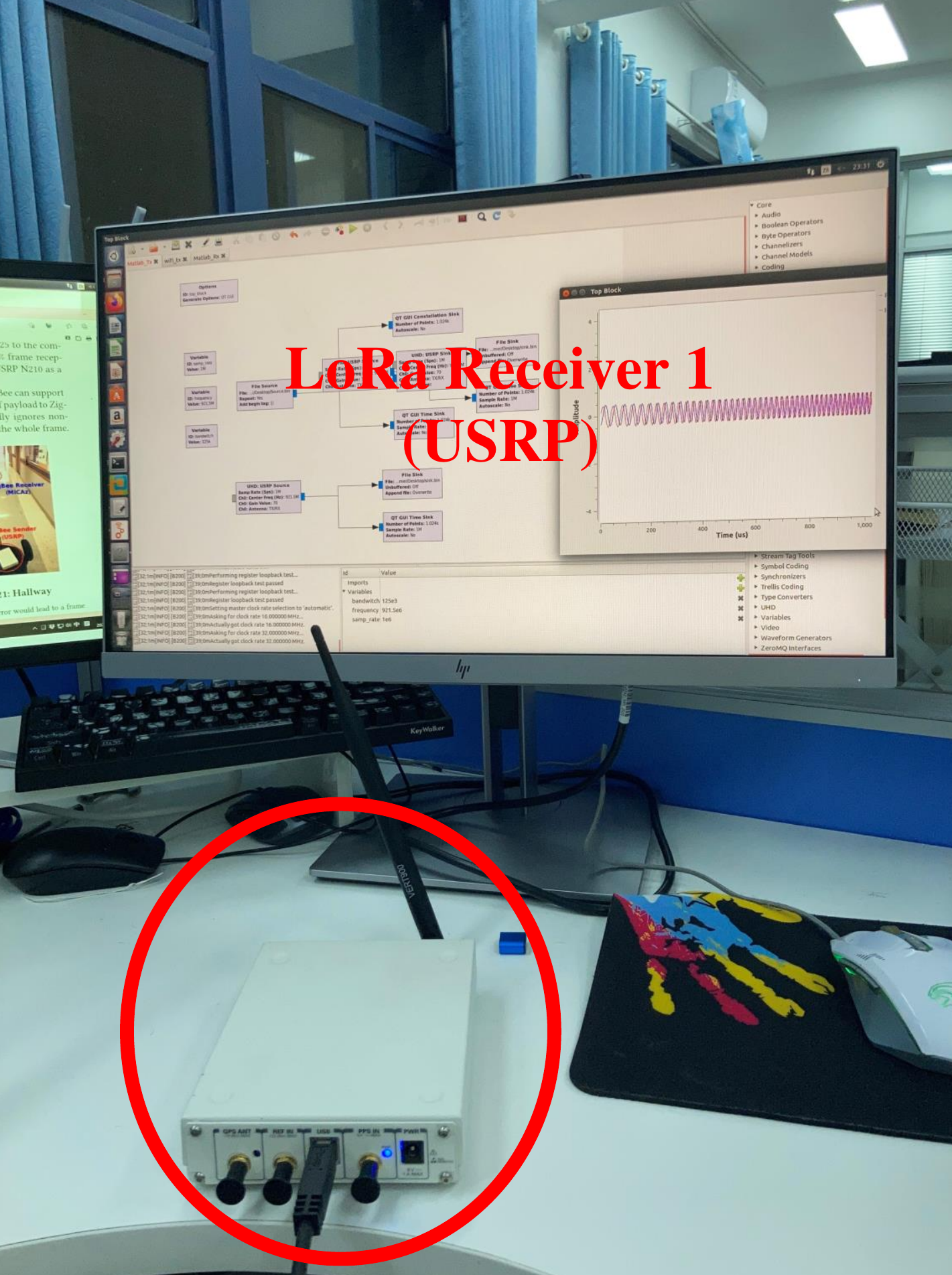}
        \caption{\textbf{Lab}}
        \label{fig:side:a}
    \end{minipage}%
    \begin{minipage}[t]{0.2\linewidth}
        \centering
        \includegraphics[width=2.3cm]{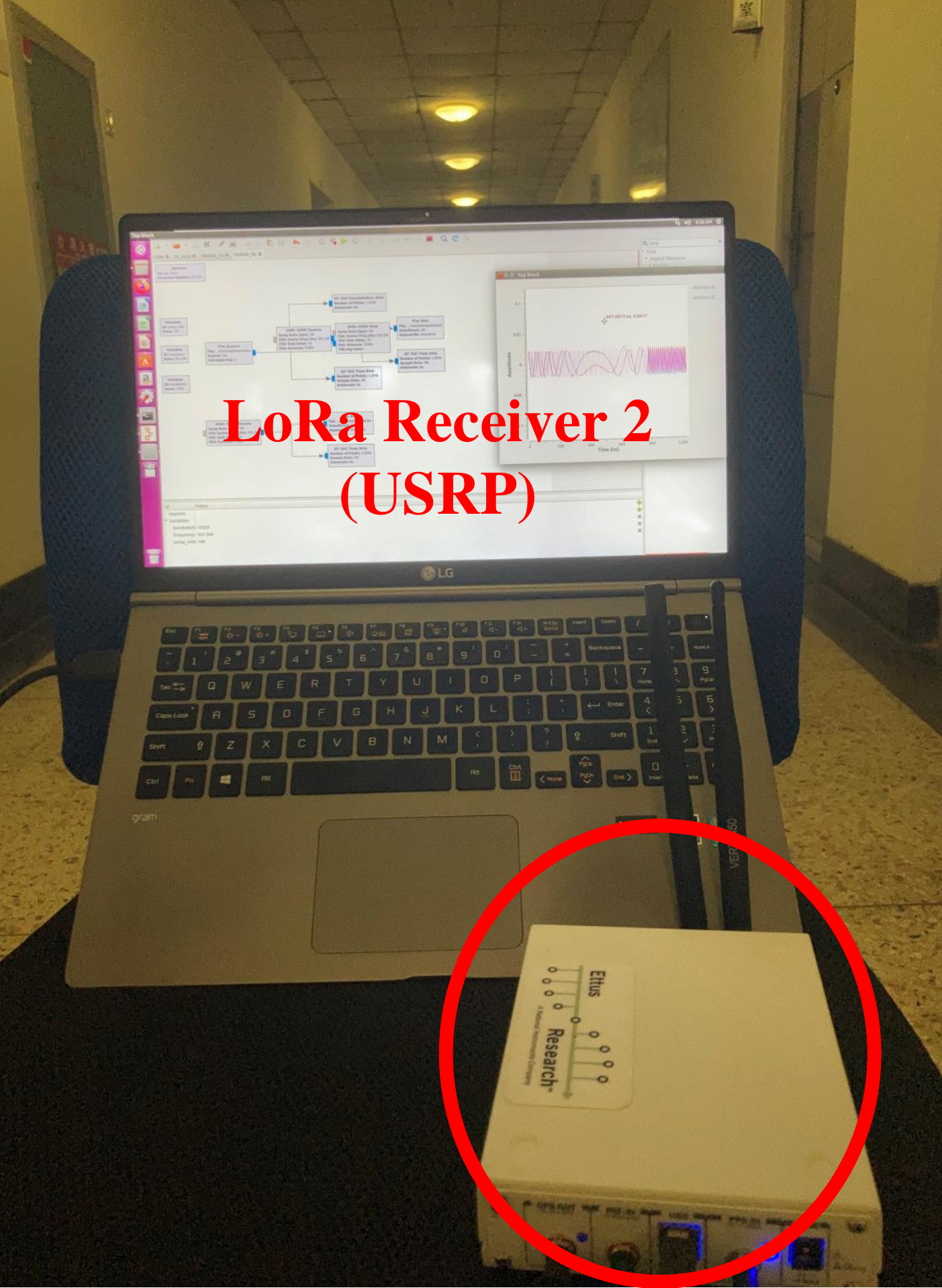}
        \caption{\textbf{Hallway}}
        \label{fig:side:b}
    \end{minipage}%
    \begin{minipage}[t]{0.25\linewidth}
        \centering
        \includegraphics[width=3.5cm]{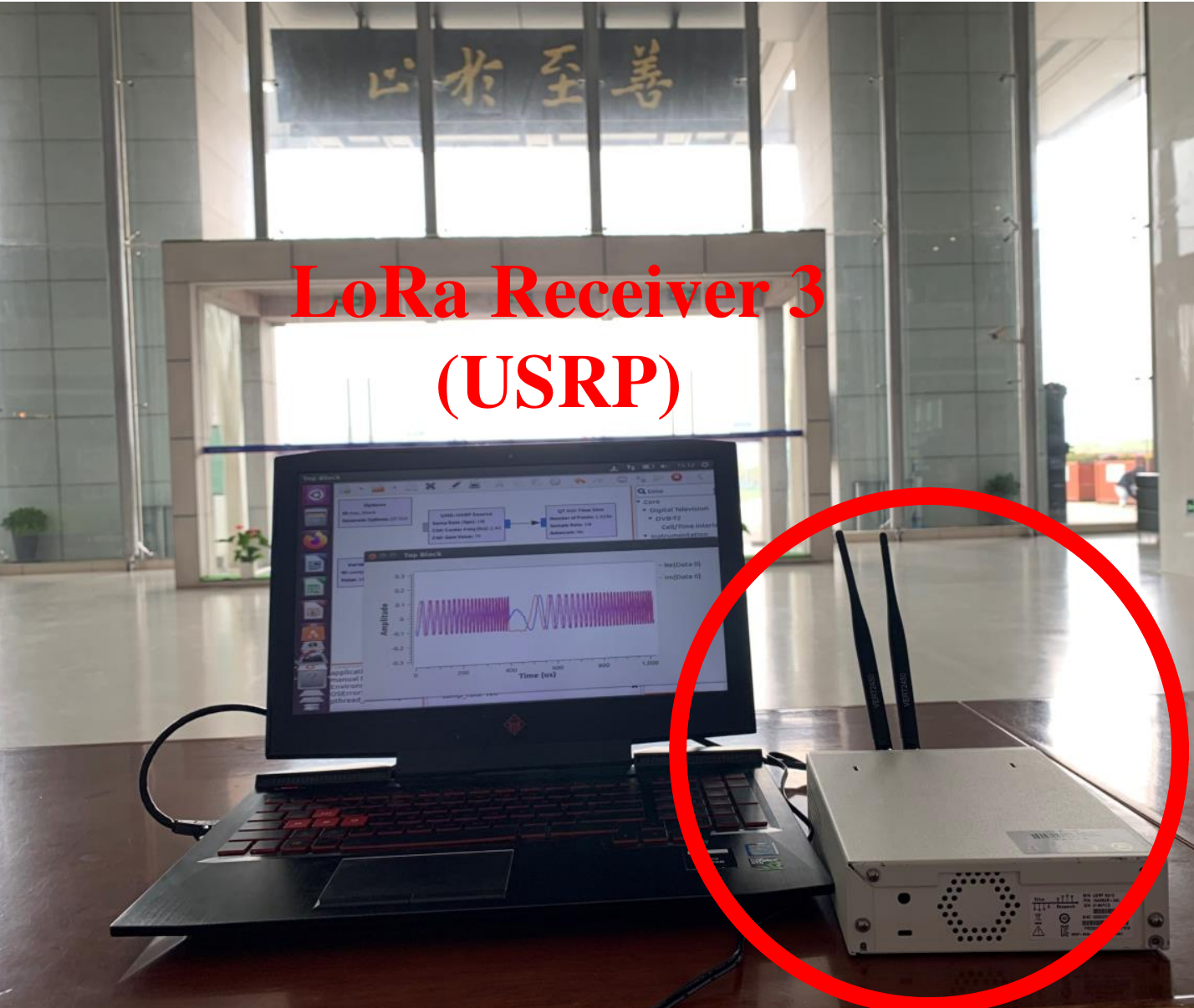}
        \caption{\textbf{Library}}
        \label{fig:side:c}
    \end{minipage}%
    \begin{minipage}[t]{0.4\linewidth}
        \centering
        \includegraphics[width=3.7cm]{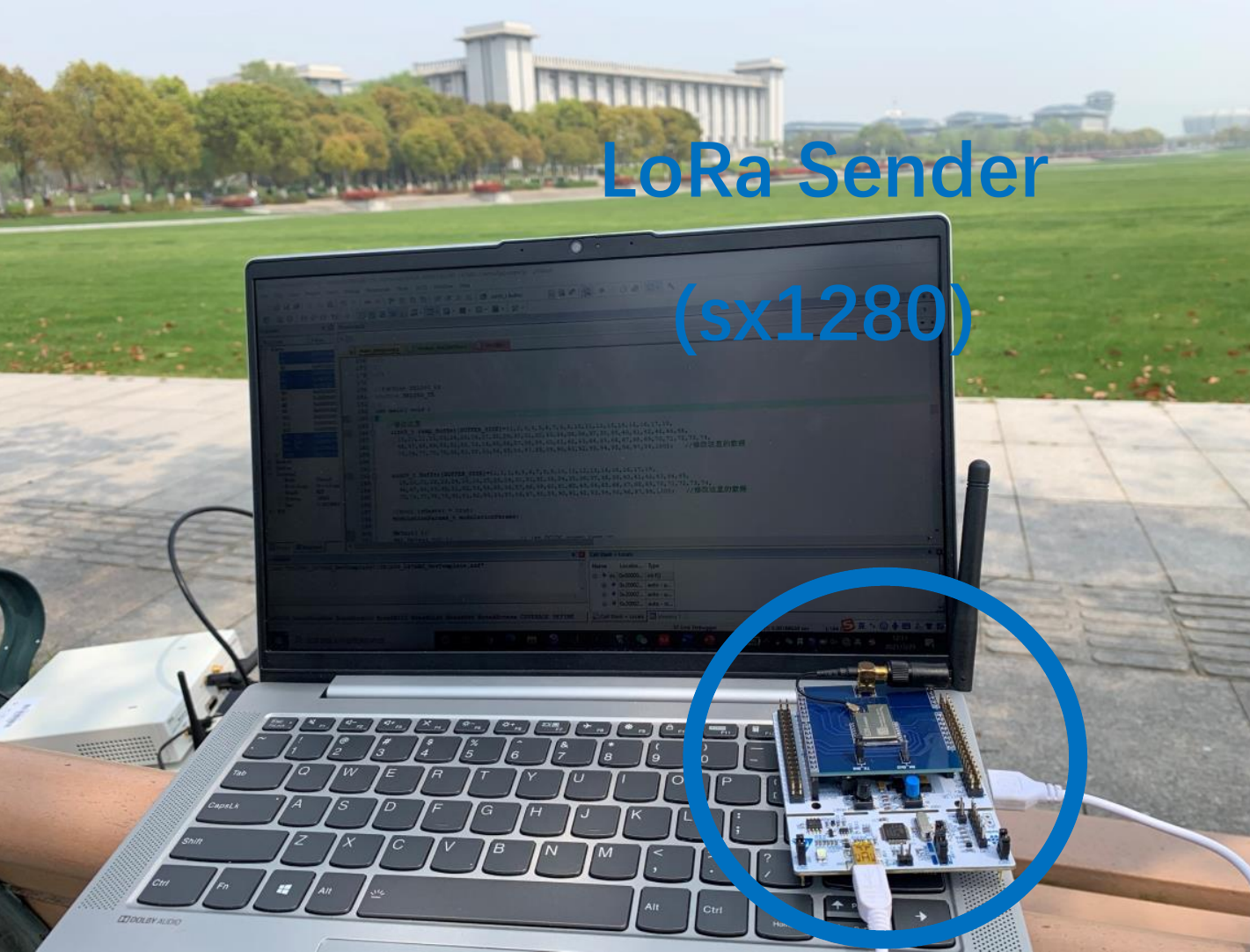}
        \caption{\textbf{Outdoor Site}}
        \label{fig:side:d}
    \end{minipage}
\end{figure}
\vspace{-0.8cm}

\vspace{-0.6cm}
\begin{figure}[H]
    \begin{minipage}[t]{0.5\linewidth}
        \centering
        \includegraphics[width=6.1cm]{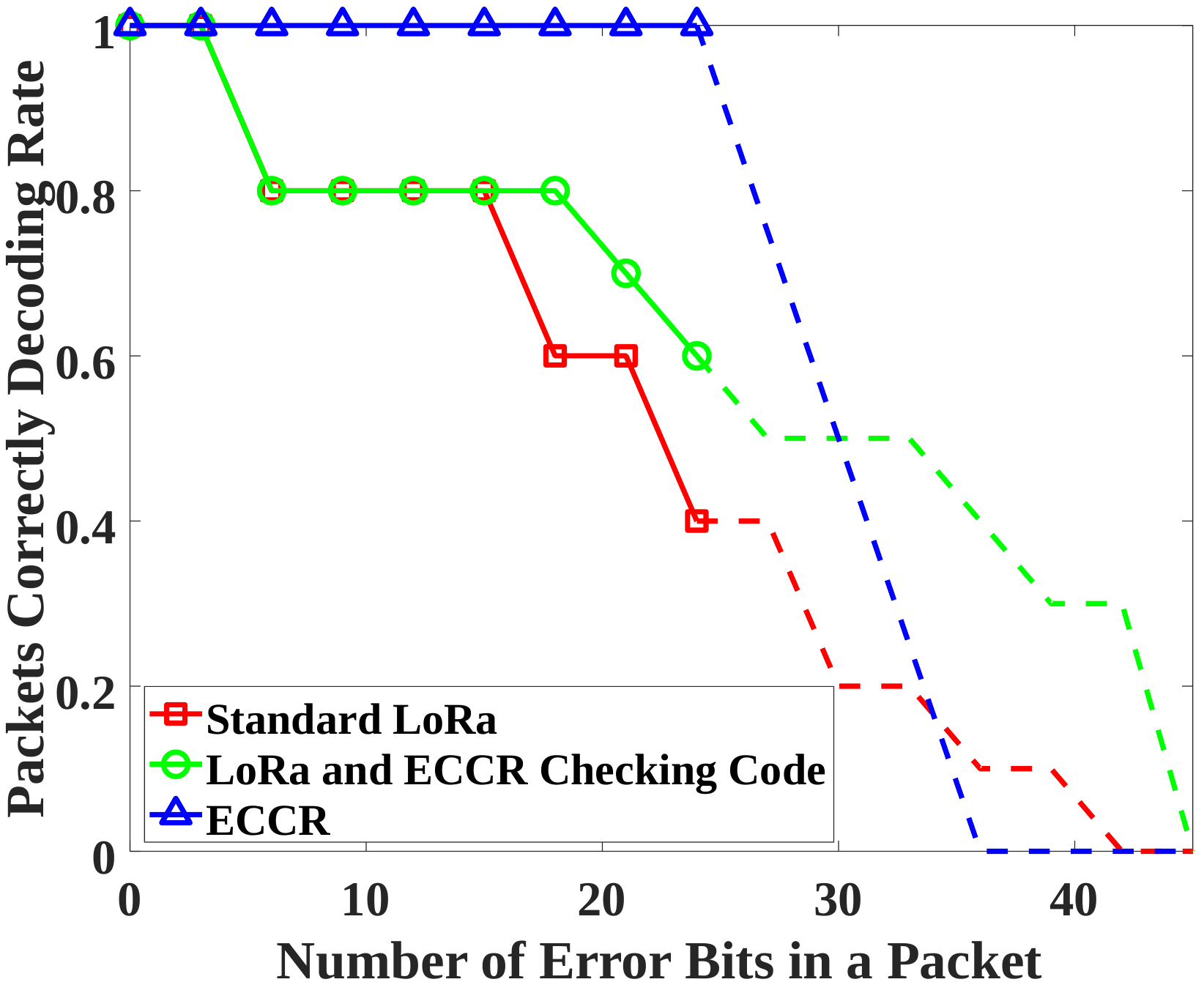}
        \caption{\textbf{Packets Decoding Rate}}
        \label{fig:side:a}
    \end{minipage}%
    \begin{minipage}[t]{0.5\linewidth}
        \centering
        \includegraphics[width=6.1cm]{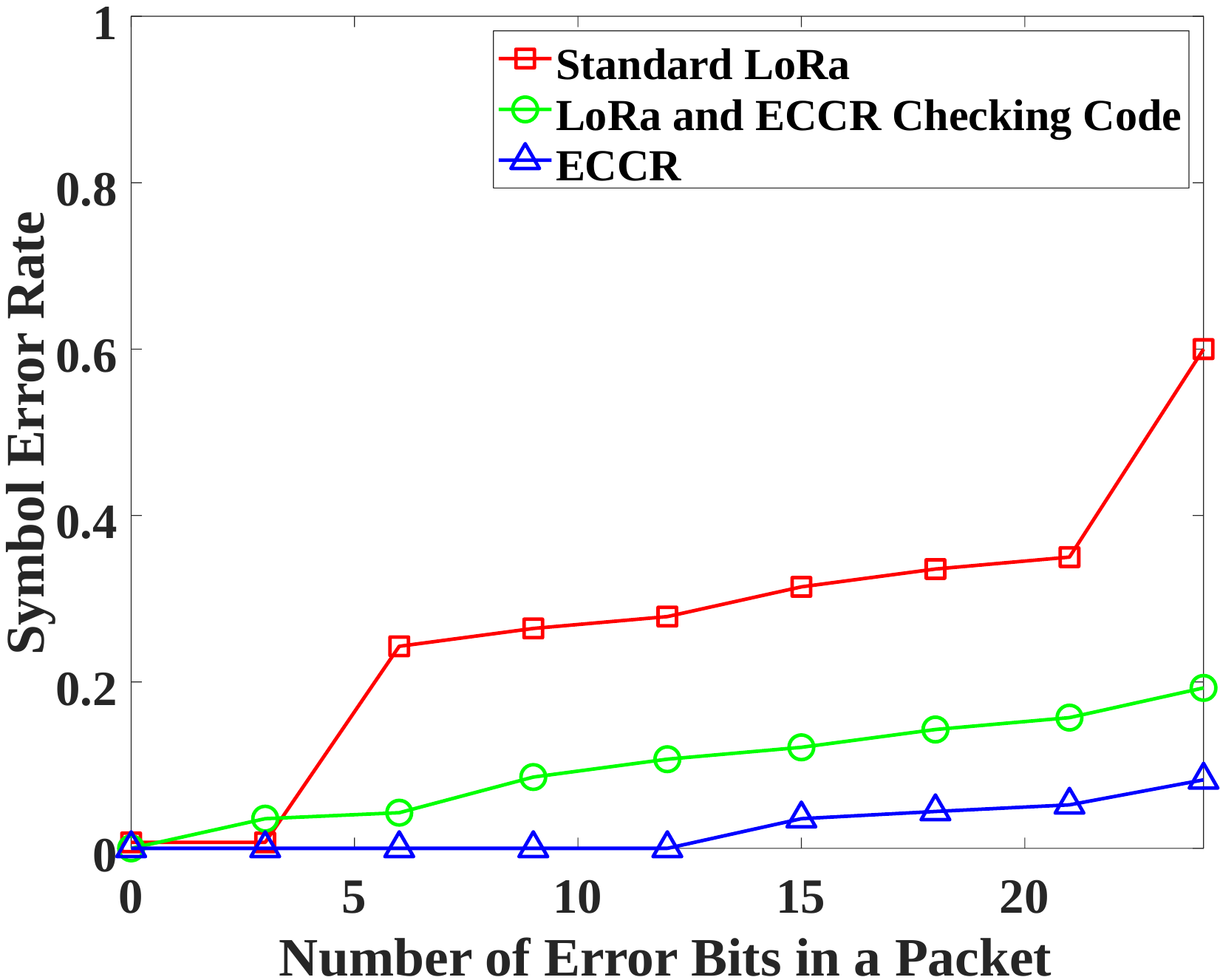}
        \caption{\textbf{Symbol Error Rate (SER)}}
        \label{fig:side:b}
    \end{minipage}
\end{figure}
\vspace{-0.4cm}

\vspace{-1.2cm}
\subsubsection{Performance of Packets Correctly Decoding Rate.} Fig. 11 shows the Packets correctly Decoding Rate of three scenarios when the interference duration increases. Packets correctly decoding rate of (i) and (ii) both decline when the interference extend. ECCR maintains $100\%$ correctly decoding rate until there are more than 30 error bits in a packet. Notice that because both (i) and (ii) utilize forward error correction code for interference mitigation the decoding rates steadily declines when increasing the duration of the interference. ECCR takes advantage of the correct information from multiple receivers so it faces a drop of correctly decoding rate when the duration of interference exceeds a boundary (30 error bits in our experiments).  
\vspace{-0.6cm}
\subsubsection{Performance of Symbol Error Rate.} Fig. 12 compares the Symbol Error Rate (SER) in three different scenarios. Notice that adding ECCR checking codes reduces SER, and ECCR achieves the lowest SER. Specifically, the error correction capability of LoRa is increased by adding ECCR checking codes (see Section 3.2 error detection). Besides, the use of weight voting algorithm further reduces SER.

Fig. 11 and Fig. 12 show that ECCR maintains accurately decodes packets when packets have $51.76\%$ corruption (in original LoRa) and it also reduces SER by $51.76\%$.

\vspace{-0.8cm}
\begin{figure}[H]
    \centering
    \includegraphics[width=11cm]{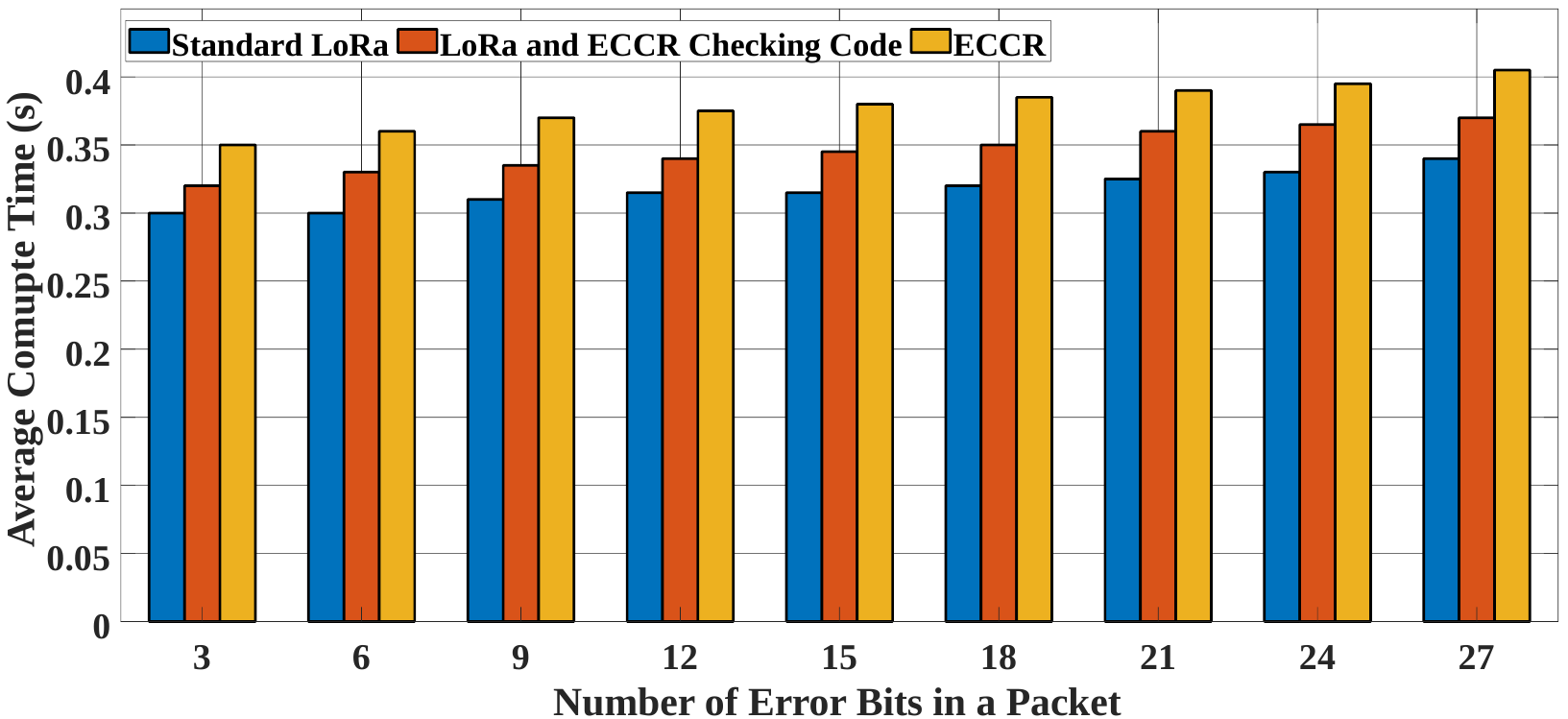}
    \caption{\textbf{Average Computation Time}}
    \label{Fig13}
\end{figure}
\vspace{-0.6cm}

\vspace{-0.8cm}
\subsubsection{Average Computation Time.} Fig. 13 shows the computing time performance of three different scenarios, where adding ECCR checking code contributes $17\%$ computation time and weight voting algorithm contributes $10\%$ computation time. ECCR, which costs $27\%$ of average computation time, achieves to accurately decode packets when packets have nearly $51.76\%$ corruption (In the strongest interference situation, $27$ milliseconds).

\vspace{-0.4cm}
\section{\uppercase{Related works}}
\vspace{-0.4cm}

This section summarizes the most related works of this paper. Most of the efforts on the LP-WANs interference mitigation fall into the following two categories. LP-WANs interference mitigation mostly on re-designing LoRa base stations and end devices \cite{ref3,ref4,ref5,ref6}, is protocol-based. Recently, using cloud computing resources for recovering corrupted packets has emerged as a mechanism for information collaboration for interference mitigation, these cloud-based methods show great compatibility to deployed LP-WANs systems.

\vspace{-0.3cm}
\subsubsection{Protocol-based Approaches.} 
\vspace{-0.3cm}
Early efforts on interference mitigation in LP-WANs have focused on solutions to physical, and MAC layers, including SCLoRa \cite{ref17}, Choir \cite{ref23}, FTrack \cite{ref24}, mLoRa \cite{ref25}, etc. in physical layer and S-MAC \cite{ref26}, LMAC \cite{ref27}, etc. in MAC layer. These protocol-based solutions require re-designing LP-WANs senders and/or base stations. The requirement on dedicated devices greatly limits the large-scale application for those approaches.

\vspace{-0.3cm}
\subsubsection{Cloud-based Approaches.}
\vspace{-0.3cm}
Benefit from the architecture of the LP-WANs system, it is feasible to utilize cloud resources for interference mitigations. For example, OPR \cite{ref2} offloads RSSI samples and the corrupted packets together send to the cloud, and utilizes the cloud to compute and restore packets. Taking advantage of the ample computational resources and the global management ability of the cloud, cloud-based approaches achieve great progress in recent researches. Besides, those approaches show great compatibility to deployed LP-WANs systems, since they don't require any hardware modification. However, offloading all the RSSI samples to the cloud incurs excessive communication overhead in the uplink bandwidth.

ECCR proposed in this paper is the first work to realize error detecting at the base sation side without transmitting RSSI sample to the cloud. It greatly reduces the data transmission amount. Also, ECCR utilizes a weight voting algorithm to collaborates correct information from multiple base stations, so that it has the capability to recovery packets with low compute complexity.

\section{\uppercase{conclusion}}
\vspace{-0.4cm}

This work presents ECCR, an edge-cloud collaborative recovery design for interference mitigation. Taking the advantage of both the global management ability of the cloud and the signal to perceive the benefit of each base stations, ECCR is the first work, to the best of our knowledge, to implement the interference mitigation based on edge-cloud collaborative recovery, which achieves to accurately decode packets when packets have nearly 50\% corruption and reduce SER for 50\%. Our experiments show that ECCR achieves correctly decoding packets when there is 50\% corruption.

In the future, we will further focus on the case with insufficient receivers (e.g., two base stations). ECCR, explores a new methodology to achieve interference mitigation with edge-cloud collaboration, and achieves a nice balance among reliability, flexibility, deployability, and complexity.

\section{\uppercase{acknowledgement}}
\vspace{-0.4cm}
This work was supported in part by National Natural Science Foundation of China under Grant No. 61902066, Natural Science Foundation of Jiangsu Province under Grant No. BK20190336, China National Key R\&D Program 2018YFB2100302 and Fundamental Research Funds for the Central Universities under Grant No. 2242021R41068


\end{document}